\documentclass{ws-procs975x65}
\begin{document}
\wstoc{Is there a quantum gravity effect on the CMB power spectrum?}{G. Esposito}

\title{Is there a quantum gravity effect on the cosmic microwave background power spectrum?}

\author{Donato Bini}
\address{Istituto per le Applicazioni del Calcolo M. Picone, CNR, 00185 Roma, Italy\\
\email{binid@icra.it}}

\author{Giampiero Esposito}
\address{INFN, Sezione di Napoli, Complesso Universitario di Monte S. Angelo,\\ 
Via Cintia, Edificio 6, 80126 Napoli, Italy\\
\email{giampiero.esposito@na.infn.it}}

\begin{abstract}
An assessment is made of recent attempts to evaluate how quantum gravity may affect the anisotropy
spectrum of the cosmic microwave background. A perturbative scheme for the solution of the
Wheeler-DeWitt equation has been found to allow for enhancement of power at large scales, whereas
the alternative predicts a suppression of power at large scales. Both effects are corrections which,
although conceptually interesting, turn out to be too small to be detected. Another scheme relies 
upon a Born-Oppenheimer analysis: by using a perturbative approach to the nonlinear ordinary
differential equation obeyed by the two-point function for scalar fluctuations, a new family of power
spectra has been obtained and studied by the authors.
\end{abstract}

\bodymatter

\vskip 0.3cm

\section{Introduction}

Since the time of its discovery \cite{Penzias}, the Cosmic Microwave Background
(hereafter CMB) radiation has provided a novel perspective on theoretical
and observational cosmology, and recent theoretical developments deal with a
careful evaluation of the CMB power spectrum.
In particular, the original idea in Ref. \refcite{Kief12} was to solve via a
JWKB method the Wheeler-DeWitt equation in the minisuperspace for a spatially
flat FRW universe coupled to a scalar field, and compute the resulting power spectrum.
The authors of Ref. \refcite{Kief12} found a suppressed CMB power spectrum at large scales.
However, in Ref. \refcite{Bini13}, the opposite result was obtained. Both papers agreed that 
the effect was too small to be detected.

Meanwhile, the work by Refs. \refcite{Kam13,Kam14}, relying upon Refs. \refcite{BV,Bertoni},
developed a Born-Oppenheimer (BO) approach to quantum cosmology that is manifestly unitary.
Within this framework, the work in Ref. \refcite{BE14} obtained new formulas for the power 
spectrum, which are assessed hereafter.

\section{Nonlinear differential equation obtained from the 
Born-Oppenheimer method}
  
The analysis is here restricted to minisuperspace for a spatially flat FRW universe
coupled to a massive scalar field. By means of a brilliant analysis, the authors of
Ref. \cite{Kam13} proved that the 2-point function ($\eta$ being conformal time)
$p(\eta)$ describing the spectrum of scalar fluctuations obeys the equation
\begin{equation}
\left[\frac{{\rm d}^{3}}{{\rm d} \eta^3}+4\omega^2\frac{{\rm d}}{{\rm d} \eta}
+2\frac{{\rm d} \omega^2}{{\rm d} \eta}\right]p+\frac{F(\eta)}{m_P^2}=0,
\label{(1)}
\end{equation}
where $p(\eta)$ pertains to the vacuum state that reduces to the
Bunch-Davies vacuum \cite{Bunch} in the short wavelength regime, 
and $F$ is found to be
\begin{eqnarray}
F(\eta) & \equiv & -{{\rm d}^{3}\over {\rm d}\eta^{3}}
\left[{({p'}^{2}+4 \omega^{2}p^{2}-1)\over 4 {a'}^{2}}\right]
+{{\rm d}^{2}\over {\rm d}\eta^{2}}\left[{p'({p'}^{2}
+4\omega^{2}p^{2}+1)\over
4p {a'}^{2}}\right] \nonumber \\
&+& {{\rm d}\over {\rm d}\eta} \left \{{1\over 8 {a'}^{2}p^{2}}
\left[(1-4 \omega^{2}p^{2})^{2}+2{p'}^{2}(1+4 \omega^{2}p^{2})
+{p'}^{4}\right]\right \} \nonumber \\
&-& {\omega \omega'({p'}^{2}+4 \omega^{2}p^{2}-1)\over
{a'}^{2}}.
\label{(2)}
\end{eqnarray}

\section{Solutions of the nonlinear differential equation}

We now define the variables $x \equiv -k\eta, \; 
\Omega \equiv \frac{\omega}{k}, P \equiv kp, \;
\varepsilon \equiv \frac{H^2}{m_{P}^{2} k^{3}}$, 
and look for solutions in the form
\begin{equation}
P(x)=\frac {1}{2} \left(1+\frac1{x^{2}}\right)+\varepsilon P_1 (x), \;   
\Omega^2(x)= 1-\frac2{x^{2}} +\varepsilon W_1 (x).
\label{(3)}
\end{equation}
By defining $V_{1} \equiv W_{1}-1$, the equation for $P_{1}$ reduces to \cite{BE14}
\begin{equation}
-P_{1}'''-4\left(1-\frac{2}{x^2}\right)P_{1}'-\left(1+\frac{1}{x^2}\right)
V_{1}'  -\frac{8}{x^3}P_1 +\frac{4}{x^3}V_{1}=0.
\label{(4)}
\end{equation}
Various particular cases can be solved, e.g.
\begin{equation}
P_{1}=0 \Longrightarrow 
V_{1} = C_{1}\left( \frac{x^2}{(1+x^2)}\right)^{2},
\label{(5)}
\end{equation}
\begin{equation}
V_{1}=0 \Longrightarrow 
P_{1} = c_{1} (Y_{-})^2+c_{2}(Y_{+})^2+c_{3} Y_{+}Y_{-},
\label{(6)}
\end{equation}
where $Y_{+}(x) \equiv \sin(x)+{\cos(x) \over x}, \;
Y_{-}(x) \equiv \cos(x)-{\sin(x) \over x}$; there exists also
a polynomial solution for both $P_{1}$ and $V_{1}$, i.e. 
\begin{equation}
P_1=\sum_{k=0}^{n_1} A_{k} x^{k} \; 
V_{1}=\sum_{k=0}^{n_2} B_{k} x^{k}.
\label{(7)}
\end{equation}
If the ratio $\varepsilon {P_{1}\over V_{1}}$ is much smaller than $1$, the
resulting perturbation can be thrusted, and the very small values of $\varepsilon$
allow for polynomials of large degree.

\section{The power spectrum}

The basic formula for the power spectrum is very simple, i.e.
\begin{equation}
{\cal P}_{\nu}={k^{3}\over 2\pi^{2}}p.
\label{(8)}
\end{equation}
In the three cases studied before, it yields
\begin{equation}
P_{1}=0 \Longrightarrow 
{\cal P}_{\nu}={k^{2}\over 4 \pi^{2}} \left(1+\frac{1}{x^2}  \right),
\label{(9)}
\end{equation}
\begin{equation}
V_{1}=0 \Longrightarrow 
{\cal P}_{\nu}={k^{2}\over 4\pi^{2}} \left\{
1+\frac{1}{x^2} +2 \varepsilon \left[
c_{1}(Y_{-})^{2}+c_{2}(Y_{+})^{2}+c_{3}Y_{+}Y_{-}\right] \right\},
\label{(10)}
\end{equation}
the third case occurring when a second degree polynomial for both $P_{1}$ and
$V_{1}$ leads to
\begin{equation}
{\cal P}_{\nu}={k^{2}\over 4 \pi^{2}}
\left[1+\frac{1}{k^{2}\eta^{2}}
+ 2\varepsilon  
(A_{0}+A_{2}k^{2}\eta^{2})\right].
\label{(11)}
\end{equation}
The constants can be adjusted to fit the data, but the data are beyond the
sensitivity of existing devices (see, however, comments in Sect. 6). The
statistical uncertainty in the data is much larger than the correction in Ref.
\refcite{Bini13}.

\section{Higher degree polynomials; powers of $x$ and $x^{-1}$}

In Ref. \refcite{BE14}, we have obtained a sixth-degree polynomial for $P_{1}$ and
$V_{1}$, with power spectrum
\begin{equation}
{\cal P}_{\nu}={k^{2}\over 4\pi^{2}}\left[1+\frac1{x^{2}} 
+2\varepsilon  P_{1}^{(6)} (x)\right].
\label{(12)}
\end{equation}
One can also consider solutions which involve both negative and positive powers of $x$.
Interestingly, a particular solution exists for which
\begin{equation}
P_{1}= -{1\over 4}{B_{-3}\over x^{3}}
-{1\over 2}{B_{0}\over x^{2}},
\label{(13)}
\end{equation}
\begin{equation}
V_{1}={B_{-3}\over x^{3}}-{B_{1}\over x}+B_{0}.
\label{(14)}
\end{equation}
The resulting power spectrum is
\begin{equation}
{\cal P}_{\nu}={k^{2}\over 4 \pi^{2}}\left[1+\frac1{x^{2}} 
-{\varepsilon \over x^{2}}\left({B_{-3}\over x}+B_{0}\right)\right],
\label{(15)}
\end{equation}
which has therefore an enhancement factor equal to 
$-{(B_{-3}+B_{0}x)\over x(x^{2}+1)}$, whose average is equal to $\pi B_{0}$,
if defined as a principal-value integral \cite{BE14}.

\section{Results and new perspectives}

(i) Our formulas for the power spectrum reduce to existing ones in particular cases, but
have better potentialities, because the lower and upper limit of summation for $P_{1}$
and $V_{1}$ are arbitrary.
\vskip 0.3cm
\noindent
(ii) The work in Ref. \refcite{Kam14} has evaluated the spectra of scalar and tensor perturbations
to first order in the slow-roll approximation, which provides new quantum gravity effects 
with respect to pure de Sitter.
\vskip 0.3cm
\noindent
(iii) The work in Ref. \refcite{Kam15} has found that the effect of quantum gravitational
corrections compared to the standard cases leads to a better fit with the data.
The possibility of understanding the loss in power as an effect of quantum 
gravity on standard slow-roll inflation is an entirely new perspective, 
and the results by these authors suggest looking for
quantum gravity effects at much smaller scales as well. 

\section*{Acknowledgments}
G. Esposito is grateful to Dipartimento di Fisica of Federico II University, Naples,
for hospitality and support. The authors have enjoyed previous collaboration with
C. Kiefer, M. Kr\"{a}mer and F. Pessina on related topics, and thank A. Yu. Kamenshchik
for correspondence.

\vfill

\end{document}